# The Gibbs paradox

Quanmin Guo

*School of Physics and Astronomy, University of Birmingham, Birmingham B15 2TT, UK*

**Abstract:** Molecular collision within an ideal gas originates from an intrinsic short-range repulsive interaction. The collision reduces the average accessible physical space for a single molecule and this has a direct consequence on the entropy of the gas. The accessibility of a molecule to a spatial coordinate (x, y, z) inside the system depends on the local molecular density. By considering mechanical equilibrium between a system and a reservoir, the probability of the system in state *i* with volume $v_i$ is shown to be proportional to $e^{-\frac{v_i}{v_0}}$ where $v_0$ is the average volume per molecule. Incorporating this factor into the single particle partition function automatically leads to an *N*-particle entropy that is extensive without applying the *N!* correction factor. The $e^{-\frac{v_i}{v_0}}$ factor plays a similar role in describing the volume distribution as the Boltzmann factor which governs the energy distribution.

An ideal gas is generally described as a gas in which molecules do not interact with each other. Strictly speaking, molecules within an ideal gas do interact with each other via molecular collisions. The molecular collision originates from a short-range intermolecular repulsion which so far has not been described with any mathematical expression. An ideal gas is thus different from an absolute non-interacting gas. For an absolute non-interacting gas, the trajectory of a molecule is not affected by other molecules in the system. For an ideal gas, the trajectory is heavily dependent on other molecules. The collision between molecules also facilitates energy exchange between molecules, and more importantly it plays an important role in the entropy of the gas. If an ideal gas is treated as absolute non-interacting, applying standard statistical analysis would lead to over-counting of states and non-extensive entropy. Here I derive the entropy for an ideal gas by considering the intermolecular repulsive interaction and demonstrate that entropy of the gas is extensive without the need of applying the *N!* factor nor the argument about indistinguishability.

For an ideal gas consisting of *N* molecules, the entropy difference between two macrostates ($T_i$, $V_i$) and ($T_f$, $V_f$) can be directly evaluated from the differential form of the first law of thermodynamics, $dU = TdS - pdV$. Simple integration gives:

$$\Delta S = \frac{3}{2} Nk \ln \frac{T_f}{T_i} + Nk \ln \frac{V_f}{V_i} . \qquad (1)$$

The entropy of the *N* molecule system takes the following form:

$$S = \frac{3}{2} Nk \ln T + Nk \ln V + C , \qquad (2)$$

where *k* is the Boltzmann constant. Although *C* in Eq. (2) is often treated as a constant when the particle number is fixed, it is more than just a constant. As discussed by Jaynes [1], the accurate form of entropy, when considering its dependence on the particle number, is:

$$S(T,V,N) = \frac{3}{2} Nk \ln T + Nk \ln V + kf(N) , \qquad (3)$$

where $f(N)$ is a function of $N$. Applying the condition that entropy must be extensive, Pauli found a general solution of $f(N)$ [2]:

$$f(N) = Nf(1) - N\ln N, \qquad (4)$$

where $f(1)$ is a constant. Hence, the entropy expressed as a function of $T$, $V$ and $N$ is:

$$S(T,V,N) = Nk\left[\frac{3}{2}\ln T + \ln\frac{V}{N} + f(1)\right]. \qquad (5)$$

Eq. (5) shows that the entropy depends on $T$ and $V/N$ where $V/N$ is the average volume per particle. For a closed system with constant $N$, Eq. (5) reduces to Eq. (2). In this article, I will discuss how the partition function depends on particle density and explore the possibility of deriving Eq. (5) directly from the partition function of the system, and further discuss the origin of the Gibbs paradox.

In the standard statistical mechanical treatment of canonical ensembles, the entropy of an ideal gas system in thermal equilibrium at temperature $T$ can be directly derived from the partition function $Z(T, V, N)$. For $N$ independent molecules, $Z(T,V,N) = Z_1^N$ where $Z_1$ is the single molecule partition function. However, the entropy derived from such a partition function for an ideal gas has the well-known problem of being non-extensive and it leads to the Gibbs paradox [3]. Many attempts have been made to rectify this problem and the popular solution found in many textbooks [4] is to use $Z(T,V,N) = Z_1^N \times \frac{1}{N!}$.

The introduction of $\frac{1}{N!}$ is claimed to be necessary for systems containing indistinguishable particles to remove over-counted states. The legitimacy of the $\frac{1}{N!}$ factor together with the argument around distinguishability of particles has been subject to intensive debate up to the present day [5-11]. Here I show that over-counting of states occurs in the initial step where $Z_1$ is evaluated, rather than due to the issue with distinguishability.

I will begin by considering how $Z_1$ is evaluated following the textbook convention. The partition function for a system of volume $V$ containing just a single molecule is given by:

$$Z_1 = \frac{1}{h^3} \int d^3q \, d^3p \, e^{-\frac{p^2}{2mkT}}. \tag{6}$$

In the above equation, $q$ is the position and $p$ the momentum of the molecule, respectively. $h$ is the Planck constant. The integral over position equals the volume of the system:

$$\int d^3q = V.$$

Moreover,

$$\begin{aligned}\int d^3p \, e^{-\frac{p^2}{2mkT}} &= \int d^3p \, e^{-\frac{p_x^2+p_y^2+p_z^2}{2mkT}} \\ &= \int dp_x e^{-\frac{p_x^2}{2mkT}} \int dp_y e^{-\frac{p_y^2}{2mkT}} \int dp_z e^{-\frac{p_z^2}{2mkT}}\end{aligned} \tag{7}$$

The partition function for a single molecule is thus:

$$Z_1 = V \left(\frac{2\pi mkT}{h^2}\right)^{\frac{3}{2}} = \left(\frac{V}{\lambda^3}\right), \tag{8}$$

where $\lambda = \sqrt{\frac{h^2}{2\pi mkT}}$ is the thermal wavelength.

The next step becomes crucial when we consider the partition function for a system containing $N$ molecules inside volume $V$. Following the textbook approach, the partition function for $N$ independent molecules within the same volume is:

$$Z_N = Z_1^N \tag{9}$$

Eq. (9) is written based on the assumption that an arbitrary particle in the system is completely independent of the presence of other particles in the system. The question is how independent is a particle co-existing with other particles in the same volume $V$? For ideal gas molecules, there is not an explicit expression describing the inter-molecule potential. However, it is always understood that a short-range repulsive interaction exists. This repulsive interaction is responsible for the chemical potential as well as the pressure of the system. If both $V$ and $T$ are fixed, adding particles to the system increases its chemical potential. Since the chemical potential is an intensive quantity, the change to the chemical potential is direct evidence that the property of the individual molecule in the system depends on $N$. The chemical potential is a function of state, so is the partition function. If the chemical potential changes with $N$, so must be the partition function. Eq. (8) is the partition function for a single molecule in volume $V$ without any other molecules. If there are $N$ molecules sharing the same volume $V$, $Z_1$ should be a function of $N$ and written as $Z_1(N)$.

Lets examine the system with only one molecule in volume $V$ in thermal equilibrium with a large reservoir at temperature $T$ as shown in Figure 1(a). The partition function for the system is given by Eq. (8):

$$Z_1 = V\left(\frac{2\pi mkT}{h^2}\right)^{\frac{3}{2}} = \left(\frac{V}{\lambda^3}\right)$$

We now create $N$ identical single molecule systems and put them all in thermal contact at $T$, Fig. 1(b). Apart from thermal energy exchange, the $N$ systems are completely independent of each other. The partition function for the $N$ systems as a whole is thus:

$$Z_N = Z_1^N = \left[\frac{V^N}{\lambda^{3N}}\right]$$
(10)

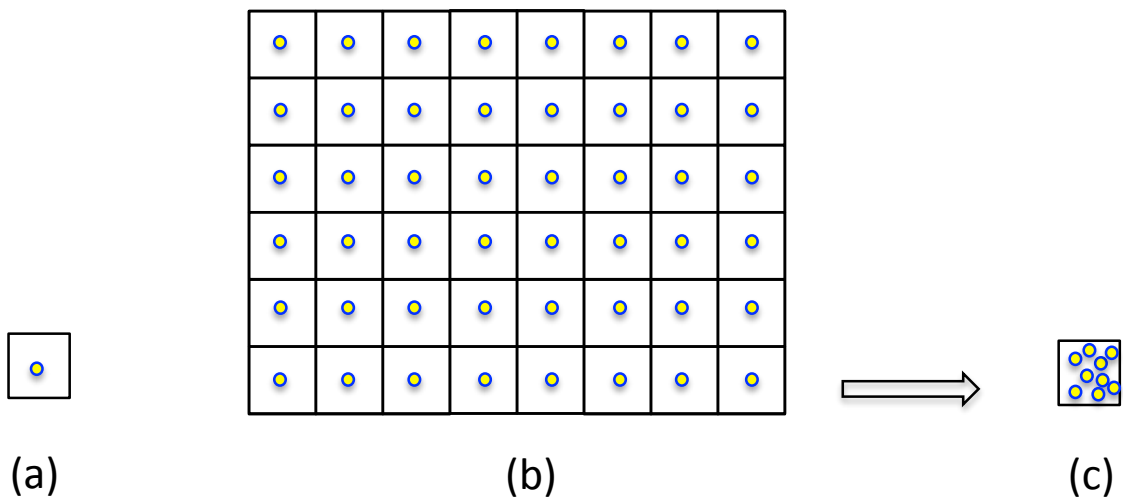

(a)          (b)          (c)

Figure 1. (a) A single molecule system of volume V at temperature T. (b) N identical single molecule systems all under thermal equilibrium. (c) A N-molecule system of volume V at temperature T.

This leads directly to the entropy of the $N$ systems:

$$S = Nk\left[\frac{3}{2} + \ln\left(\frac{V}{\lambda^3}\right)\right]$$
$$= Nk\left[\frac{3}{2} + \ln V + \frac{3}{2}\ln\left(\frac{2\pi mkT}{h^2}\right)\right] \quad (11)$$

Figure 1(c) shows a single system with volume $V$ containing $N$ molecules at $T$. We can conduct a reversible process to transform (b) into (c) and hence evaluate the entropy for (c). We first remove the walls between neighboring systems in (b) to allow the $N$ systems to be connected. We then perform an isothermal compression to reduce the volume from $NV$ to $V$. The final product is system C. The isothermal compression leads to a reduction of entropy of the system by $Nk\ln N$. Therefore, the entropy of system (c) is:

$$S = Nk\left[\frac{3}{2} + \ln V + \frac{3}{2}\ln\left(\frac{2\pi mkT}{h^2}\right)\right] - Nk\ln N = Nk\left[\frac{3}{2} + \ln\left(\frac{V}{N}\right) + \frac{3}{2}\ln\left(\frac{2\pi mkT}{h^2}\right)\right] \quad (12)$$

Eq. (12) is the same as Eq. (5) with

$$f(1) = \frac{3}{2}\left[1 + \ln\left(\frac{2\pi mk}{h^2}\right)\right] \quad (13)$$

In transforming (b) to (c), the first step where the walls of the sub-systems are removed needs some discussion. From a classical thermodynamics point of view, the removal of walls is a reversible process involving neither heat exchange between the systems and the reservoir nor work performed onto the systems. Thus, there is no change to entropy to either the system or the reservoir when the walls are removed.

The entropy for the *N* molecule system, Fig. 1(c), can be derived from its partition function. In Fig. 1(c), the total volume is shared among *N* molecules. The repulsive interaction between molecules gives rise to molecular collision and the molecule has a mean free path. The mean free path of the molecule can be used to evaluate the available physical space around each molecule. For instance, the mean free path of a molecule in atmosphere at room temperature is only around 60 nm with the average nearest neighbor distance of molecules around the same value. Thus, the physical space available to a single molecule in atmosphere is ~ $10^{-22}$ m$^3$. A single molecule does not travel much further than its mean free path before a head on collision. These frequent collision events set up a physical boundary around each molecule limiting its footprint to a small volume. The footprint of the molecule diffuses within the total volume *V* with a speed much slower than the molecular speed. Thus, each molecule, on average, carries about *V/N* "specific space" as it diffuses in the system. A molecule keeps changing its trajectory because its way of travel is blocked by other molecules making the vast quantity of the volume inaccessible. The fact that given a long enough time, a molecule would have eventually sampled all the physical space provided by V is a different issue which will be discussed later. In Eq. (6), the Boltzmann factor makes a molecule with exceedingly high energy very improbable. In the following, I will show that a similar factor should be incorporated to make a molecule having exceedingly large specific volume also improbable.

We consider a system under equilibrium with a large reservoir as shown in Fig. 2. The system can exchange heat but not particles with the reservoir. The walls of the system are flexible allowing free volume exchanges between the system and the reservoir. The total volume is $V_T$ and the total energy is $E_T$.

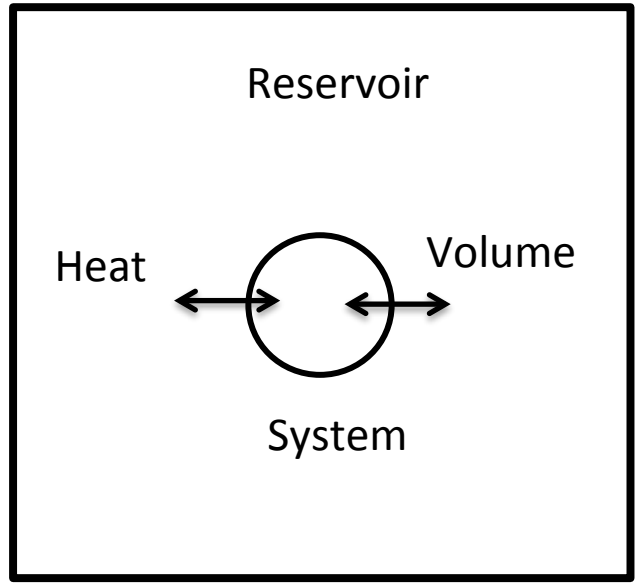

Figure 2. A system in equilibrium with a reservoir at temperature T and pressure P. The system does not exchange particles with the reservoir, but it can exchange volume by expanding into the reservoir or vice versa. The wall of the system is flexible so it does not add any extra energy during expansion or contraction.

The probability, $P_i$, of finding the system in state $i$ (a specific microstate), having energy $E_i$ and volume $v_i$ is:

$P_i = \frac{Number\ of\ states\ of\ the\ bath\ corresponding\ to\ system\ in\ state\ i(E_i,v_i)}{Total\ No.states\ for\ system\ and\ bath\ combined(including\ all\ possible\ E_i\ and\ v_i)}.$

$$P_i \propto \Omega_{Total} = \Omega_{bath}(E_T - E_i, V_T - v_i), \quad (14)$$

$$\Omega_{bath}(E_T - E_i, V_T - v_i) = e^{S_{bath}/k}. \quad (15)$$

Perform Taylor expansion for $S_{bath}$.

$$S_{bath}(E_T - E_i; V_T - v_i) = S_{bath}(E_T, V_T) + \frac{\partial S_{bath}}{\partial E}(-E_i) + (higher\ order\ terms) + \frac{\partial S_{bath}}{\partial V}(-v_i) + (higher\ order\ terms). \quad (15)$$

We know that, $\frac{\partial S_{bath}}{\partial E} = \frac{1}{T}$, $\frac{\partial S_{bath}}{\partial V} = \frac{P}{T}$, $T$ and $P$ are constants, thus higher order terms are zero.

$$S_{bath}(E_T - E_i, V_T - v_i) = S_{bath}(E_T, V_T) - \frac{1}{T}E_i - \frac{P}{T}v_i. \quad (16)$$

$$P_i \propto e^{S_{bath}(E_T, V_T) - \frac{E_i}{T} - \frac{P}{T}v_i}/k = \alpha e^{(-E_i - Pv_i)/kT}. \tag{17}$$

Substituting $PV_T = N_T kT$ into (17):

$$P_i \propto e^{\left(-\frac{E_i}{kT}\right)} e^{\left(-\frac{v_i}{V/N}\right)} = e^{\left(-\frac{E_i}{kT}\right)} e^{\left(-\frac{v_i}{v_0}\right)}, \tag{18}$$

where $v_0 = \frac{V_T}{N_T}$ is the average volume per molecule. If the volume of the system is held constant, then Eq. (18) reduces to the simple Boltzmann factor.

For the system shown in Fig. 1(c), we can treat it as composed of $N$ subsystems, each of which has one single molecule. Each subsystem can exchange energy as well as volume with other subsystems, but not particles. For example, the molecule in an arbitrary subsystem can gain or lose volume via its collisions with neighbouring molecules. As a molecule moves through the sea of other molecules, the inherent volume attached to the molecule changes due to fluctuation. We now apply Eq. (18) to the single molecule partition function by including the $e^{-\frac{v_i}{v_0}}$ factor.

$$Z_1 = \frac{1}{h^3} \int e^{-\frac{v_i}{v_0}} d^3q \, d^3p \, e^{-\frac{p^2}{2mkT}} = \frac{1}{h^3} \int e^{-\frac{v_i}{v_0}} dv_i \, d^3p \, e^{-\frac{p^2}{2mkT}} \tag{19}$$

The spatial part of the integration is from 0 to V.

$$Z_1 = \frac{1}{h^3} \int e^{-\frac{v_i}{v_0}} d^3q \, d^3p \, e^{-\frac{p^2}{2mkT}} = \frac{1}{h^3} v_0 \left[1 - e^{-N_T}\right] \int d^3p \, e^{-\frac{p^2}{2mkT}} = \frac{V}{N}\left(\frac{2\pi mkT}{h^2}\right)^{\frac{3}{2}} \tag{20}$$

Or,

$$Z_1(N) = \frac{V}{N}\left(\frac{2\pi mkT}{h^2}\right)^{\frac{3}{2}} \tag{21}$$

to clearly indicate that the single molecule partition function is a function of $N$. In the last step of Eq. (20), we have applied the condition that $N_T \gg 1$ which is true for standard gas systems where $N_T$ is $\sim 10^{23}$. The partition function for the whole system with $N$ molecules becomes:

$$Z_N = [Z_1(N)]^N = \left[\frac{V^N}{N^N \lambda^{3N}}\right] \qquad (22)$$

Eq. (22) leads to the same expression as Eq. (12). The entropy calculated from Eq. (22) is extensive. At no point do we need to add the $N!$ factor. We did not even need to consider the particles are distinguishable or indistinguishable. The particles are treated as identical to each other and they are all statistically equivalent.

The key step in our approach is the inclusion of $e^{-\frac{vi}{vo}}$ for the evaluation of $Z_1$. We call $e^{-\frac{vi}{vo}}$ the volume factor. The physical meaning of the volume factor is that the probability of finding a molecule at one location depends on the local particle density. This is demanded by the condition that the system has a well-defined pressure. The textbook expression of Eq. (6) assumes that the probability of finding a molecule in an arbitrary location is independent of the locations of other molecules, and this incorrect assumption leads to overestimated $Z_1$. The textbook approach of adding a factor $N!$ is explained as necessary to remove states due to permutation of particles. In our analysis, we do not discount the states from particle permutation. In classical thermodynamics, swapping two particles leads to a new microstate. As the molecular density increases in the gas system, the specific volume belonging to each molecule decreases. The molecules always manage to "fairly" share the total volume. This volume sharing process gives the V/N dependence of the single molecule partition function.

Using Eq. (21) as the single molecule partition function, the entropy for the system containing $N$ molecules comes out as an extensive quantity. If Eq. (6) is used to evaluate the single molecule partition function, the partition function becomes over evaluated. For example, Eq. (6) assumes that the following two microstates have the same probability. i) $N$ molecule uniformly distributed inside $V$; ii) a single molecule takes over *V* with the rest of *N-1* molecules occupying nearly zero volume. Qualitatively, ii) has a much lower probability. In fact, microstate ii) does not even belong to the macrostate (U, V, N) because such a state has an undefined pressure.

A molecule in an ideal gas system is not localized. It is constantly on the move. This gives an impression that the molecule can indeed access every point inside *V*. However, it is the accessible space at one instantaneous in time that is important, not the time integrated accessible space. When the partition function is evaluated, the integration is performed over the phase space with no time *t* involved. For a very large system, it may well take several days or even years for a molecule to travel from one edge of the system to the opposite edge, the non-time-integrated physical space accessible to such a molecule is given by the average volume per molecule. The rest of the physical space is populated by other molecules and this is true at all times.

We can expand our discussion to the issue of mixing of gasses. When two volumes of the same ideal gasses under the same *T* and *P* are mixed, the entropy of mixing is zero. The processing of mixing involves the exchange of molecules without altering the macrostate. Exchange of molecules does not change *V/N* and hence no change to entropy. In many textbooks, mixing is treated as expansion of two gasses into larger volumes. It needs to be clarified that mixing involves no expansion. Under steady state, molecules sharing one volume keep exchanging. The microstate of the system is changing all the time. Thus, the microscopic process is forever irreversible while the microstate remains the same.

In summary, the probability of finding a molecule at a particular location inside the volume of the system is proportional to $e^{-\frac{v_i}{v_o}}$ where $v_o$ is the average volume per molecule and $v_i$ is the volume attached to the molecule. This $e^{-\frac{v_i}{v_o}}$ factor controls how the total volume is distributed among the *N* molecules in the same way the Boltzmann factor controls how the total energy is distributed. Including the $e^{-\frac{v_i}{v_o}}$ factor for the evaluation of the single molecule partition function $Z_1$ gives rise to the entropy of the total N-molecule system to be extensive. Whether we consider the molecules as distinguishable or not does not make a difference to the entropy. The textbook assumption that a single molecule has equal probability to access any physical location inside *V*, regardless the distribution of other molecules, thus appears to be incorrect, and this incorrect assumption is the origin of non-extensive entropy discussed in the context of the Gibbs paradox.